\documentclass[12pt]{article}
\usepackage{amsmath,amssymb,yfonts,esint}
\usepackage{wrapfig}
\usepackage{floatflt}
\usepackage{mathrsfs}

\usepackage[pdftex]{color,graphicx}

\usepackage{dsfont}
\newcommand{\be}{\begin{equation}}
\newcommand{\ba}{\begin{eqnarray}}
\newcommand{\ea}{\end{eqnarray}}
\newcommand{\nn}{\nonumber}

\def\d{\delta}
\def\e{\epsilon}

\def\k{\kappa}

\def\t{\tau}

\def\G{\Gamma}

\def\S{\Sigma}

\def\ca{{\cal A}}
\def\cb{{\cal B}}

\def\ch{{\cal H}}

\def\cl{{\cal L}}

\def\cn{{\cal N}}

\def\cp{{\cal P}}

\def\cu{{\cal U}}

\newcommand{\pa}{\partial}

\newcommand{\bbC}{{\Bbb C}}

\fontfamily{yfrak}

\begin{document}

\vskip 15mm

\begin{center}

{\Large\bfseries On a Derivation of the Dirac Hamiltonian\\[1ex] From a Construction of Quantum Gravity
%\\[2mm]
}

\vskip 4ex

Johannes \textsc{Aastrup}$\,^{a}$\footnote{email: \texttt{aastrup@uni-math.gwdg.de}},
Jesper M\o ller \textsc{Grimstrup}\,$^{b}$\footnote{email: \texttt{grimstrup@nbi.dk}}\\ \& Mario \textsc{Paschke}$^{c}$\footnote{email: \texttt{mario.paschke@uni-muenster.de}}

%\& Ryszard \textsc{Nest}\,$^{c}$\footnote{email: \texttt{rnest@math.ku.dk}}

\vskip 3ex  

$^{a}\,$\textit{Mathematisches Institut, Georg-August-Universit\"at G\"ottingen,\\ Bunsenstrasse 3, 
D-37073 G\"ottingen, Germany}
\\[3ex]
$^{b}\,$\textit{The Niels Bohr Institute, University of Copenhagen, \\Blegdamsvej 17, DK-2100 Copenhagen, Denmark}\\[3ex]

$^{c}\,$\textit{Mathematisches Institut der Westf\"aischen Wilhelms-Universit\"at\\ Einsteinstrasse 62, D-48149 M\"unster, Germany}
%\\[3ex]
%$^{c}$ \textit{Mathematical Institute, University of Copenhagen,\\ Universitetsparken 5, DK-2100 Copenhagen, Denmark}
\end{center}

\vskip 3ex

\begin{abstract}

The structure of the Dirac Hamiltonian in 3+1 dimensions is shown to emerge in a semi-classical approximation from a abstract spectral triple construction. The spectral triple is constructed over an algebra of holonomy loops, corresponding to a configuration space of connections, and encodes information of the kinematics of General Relativity. The emergence of the Dirac Hamiltonian follows from the observation that the algebra of loops comes with a dependency on a choice of base-point. The elimination of this dependency entails spinor fields and, in the semi-classical approximation, the structure of the Dirac Hamiltonian. 

%The latter emerges from the expectation value of the abstract Dirac type operator on states which combines semi-classical approximation with the elimination of the choice of base-point.

\end{abstract}

\newpage

\section{Introduction}

In two recent publications \cite{Aastrup:2009dy,AGNP1} certain semi-classical states in an infinite dimensional geometrical construction over a configuration space related to General Relativity were shown to render the Dirac Hamiltonian in 3+1 dimensions. The construction is a semi-finite spectral triple \cite{AGN3,AGN1,AGN2}, which encodes the kinematical part of canonical quantum gravity \cite{AGNP1}, and the configuration space is a space of connections, related to gravity through Ashtekars approach \cite{AL1}. In both papers \cite{Aastrup:2009dy,AGNP1} the semi-classical states were laboriously engineered to match exactly the Dirac Hamiltonian in the classical limit. Thus, an immediate question arises whether these states point towards some natural structure, or whether they are mere "lucky hits".

In this paper we show that the structure, which in  \cite{Aastrup:2009dy,AGNP1} entails the Dirac Hamiltonian, is a direct consequence of working with the noncommutative algebra of holonomy loops. The product between two loops in this algebra is defined by gluing at a chosen base-point. The point made in this paper is that the construction of the algebra is sensitive to the choice of base-point. If instead one considered the algebra of traced holonomy loops, then the choice of base-point would be irrelevant since the trace discards any trivial backtracking between different choices of base-point. This mechanism is absent for the untraced loops and it is the elimination of this discrepancy that leads directly to the structure of the Dirac Hamiltonian in 3+1 dimensions.

More concretely, we shall realize the change of base-point on the Hilbert
 space of states
 by a family of operators $\tilde{\cu}_l$,  corresponding to  curves $l$ that
 connect one base-point with another.
 These operators are chosen such that given a representation $\pi_p$
 corresponding
  to one choice of base-point the representation $\pi_q$ corresponding to
  another base-point is
  obtained  as
  $\pi_q =  \tilde{\cu}_l \pi_p \tilde{\cu}_l^*$, where $l$ is an appropriate
  curve.
  
 We shall then only consider states that do not show this dependency on a base-points as being
 physical. In particular, a very natural choice of physical coherent states
 is then given as
 \[    \Psi = \sum\limits_l  \tilde{\cu}_l \psi_l \phi_n^t \;, \]
 where the $\psi_l$ are arbitrary and $\phi_n^t$ denote the natural coherent
 states on $SU(2)$
 found by Hall \cite{H1,H2}. The expectation value of an abstract Dirac type operator, which comes from the spectral triple construction over the configuration space of connections, then leads to the structure of the Dirac Hamiltonian.

What we find deviates from the Dirac Hamiltonian in two points: first, the spinors, which emerge in the semi-classical limit, carry twice the number of degrees of freedom; second, the Dirac Hamiltonian which we find involves both right and left actions on these spinors, by the connection and gamma matrices respectively.\\

This paper is organized as follows: in section 2 we briefly review the construction of the algebra of holonomy loops and, in section 3, the construction of a Dirac type operator over this algebra. In section 4 we give the basic structure of canonical quantum gravity based on Ashtekar variables and state that the spectral triple construction encodes information about the kinematics of quantum gravity. Coherent states, localized over classical points in the phase space of canonical gravity, are introduced in section 5. In section 6 we then point out that the construction of the algbra is marred with a dependency on the choice of base-point and provide a remedy in the guise of a state which divides out the base-point. Then, in section 7, we show that the expectation value of the Dirac type operator, on this state combined with a coherent state gives, in the classical limit, the form of the Dirac Hamiltonian.

\section{The algebra of based holonomy loops}

\begin{figure}[t]
\begin{center}
\resizebox{!}{2cm}{
 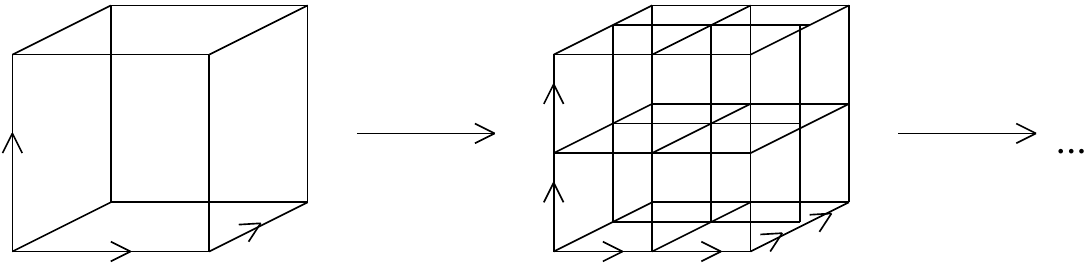}
\end{center}
\caption{\it an infinite system of nested, oriented, cubic lattices. Each lattice $\G_n$ is a symmetric subdivision of the lattice $\G_{n-1}$.}
\end{figure}

We first review the construction of the algebra of holonomy loops. For details we refer to \cite{AGN3}.
Let $\{\G_n\}_{n\in\mathbb{N}}$, be a family of nested, oriented, cubic lattices, see figure 1. One can either view these lattices in terms of a cubulation of a 3-manifold $\Sigma$ or as abstract graphs. Denote by $\{l_i\}$ the edges and by $\{x_j\}$ the vertices in the family of lattices. We choose a base-point $x_0$ in $\G_0$ and consider loops based in $x_0$. Thus, a loop $L$ is a sequence of edges $\{l_{i_1},l_{i_2},\ldots,l_{i_n}\}$ where $l_{i_1}$ starts in $x_0$ and $l_{i_n}$ ends in $x_0$. We discard trivial backtracking. Such loops have a natural noncommutative product, simply by gluing them at the base-point and a natural involution given by reversal of direction. The set of based loops in $\G_n$ generate a $\star$-algebra which we denote $\cb_{x_0}^n$. By taking the inductive limit over all lattices one obtains from $\cb_{x_0}^n$  a $\star$-algebra of loops in the infinitely refined lattice which we denote $\cb_{x_0}$. Note that this representation might not be faithful, for example if $G$ is commutative.

Given a graph $\G_n$ and a compact Lie group $G$ let $\nabla$ be a map
$$
\nabla:l_i\rightarrow g_i\in G\;,
$$
which assigns copies of $G$ to edges in $\G_n$. We denote the space of all such maps $\ca_{\G_n}$. Since
$$
\ca_{\G_n}\simeq G^{n(\G_n)}\;,
$$
where $n(\G_n)$ is the number of edges in $\G_n$, we can construct a Hilbert space over $\ca_{\G_n}$
by using the Haar measure on $G$. 

Notice that a loop $L=\{l_{i_1},l_{i_2},\ldots,l_{i_n}\}$ in $\cb^n_{x_0}$ is a natural function on $\ca_{\G_n}$ by
$$
L:\ca_{\G_n}\rightarrow G\;;\quad L(\nabla)= \nabla(l_{i_1})\cdot \nabla(l_{i_2})\cdot \ldots \nabla(l_{i_n})\;.
$$
The algebra $\cb^n_{x_0}$ can be represented as bounded operators on a Hilbert space once a matrix factor is added to accommodate the representation. %For reasons which shall be explained later 
We choose to add two matrix factors
\begin{equation}
\ch_n=L^2(\ca_{\G_n}, M_k(\mathbb{C})\oplus M_k(\mathbb{C}))\;.
\label{HIL}
\end{equation}
where $k$ is the size of a chosen matrix representation of $G$ with
$$
L\cdot \xi(\nabla) = 
\left(
\begin{array}{cc}
L(\nabla) & 0\\
0& L(\nabla)\\
\end{array}   
\right)\cdot \xi(\nabla)\;,\quad \xi\in\ch_n\;.
$$
By taking the inductive limit of hilbert spaces $\ch_n$ one obtains a limiting Hilbert space $\ch$ which carries a representation of the algebra $\cb_{x_0}$. Furthermore, if we denote by $\ca$ the space of smooth connections in a trivial bundle $\Sigma\times G$, then one can show that the projective limit of the spaces $\ca_{\G_n}$, denoted $\overline{\ca}$, contains $\ca$ as a smooth embedding \cite{AGN3}.

The entire construction works for any compact Lie group. We shall however work with $SU(2)$ since this relates the construction to canonical quantum gravity formulated in terms of Ashtekars variables. Furthermore, we choose the fundamental representation, $k=2$.

\section{A semi-finite spectral triple over $\cb_{x_0}$}

Since $\ca_{\G_n}$ is a manifold it is straight forward to write down a Dirac operator in $\ch_{n}$. Restrictions on this operator arise when one considers the inductive limit of Hilbert spaces since an operator $D$ in $\ch$ is constructed as a family of operator $D_n$ acting in $\ch_n$ compatible with embeddings between different Hilbert spaces. 

First, however, we must modify the Hilbert space $\ch_n$ to accommodate a Dirac operator
\begin{equation}
\ch_n=L^2(\ca_{\G_n}, Cl(T^\ast G^{n(\G_n)})\otimes M_k(\mathbb{C}))\;,
\label{Hilbert}
\end{equation}
where $Cl(T^\ast G^{n(\G_n)})$ is the Clifford bundle over $G^{n(\G_n)}$. At the level of a graph $\G_n$ a Dirac type operator, which is compatible with embeddings $P^\ast:\ch_n\rightarrow \ch_{n+1}$, is of the form
\begin{equation}
D_n = \sum a_i {\bf e}_i^a \cl_{{\bf e}_i^a}\;,
\label{ddd}
\end{equation}
where $ \cl_ {{\bf e}_i^a}$ denote a derivation with respect to a left-translated vector field on the $i$'th copy of $G$ and $ {\bf e}_i^a$ denote both a left-translated vector field and its corresponding element in the Clifford algebra. As elements in the Clifford algebra they are subjected to the conventions ${\bf e}_i^a {\bf e}_i^a = -1$ (no sum). Also, $\{a_i\}$ is a set of real parameters. The construction of $D_n$ involves a change of variables which means that the sum over $i$ in (\ref{ddd}) should be understood with respect to this change of variables. For details we refer to \cite{AGN3}. Once the requirement of compatibility is met $D_n$ gives rise to a densely defined operator $D$ in $\ch$.

It turns out that the collection $(\cb_{x_0},\ch,D)$ satisfy the requirements of a semi-finite spectral triple when the parameters $a_i$ approach infinity with increased subdivision of lattices. To have a spectral triple means that two conditions are satisfied: 1) the resolvent of $D$, $(1+D^2)^{-1}$, is compact and 2) the commutator $[D,a]$, with $a\in\cb_{x_0}$, is bounded. The term semi-finite means that the first condition is only partially satisfied. The resolvent of $D$ is only compact with respect to a certain trace. In the present case we find that the spectrum of $D$ is infinitely degenerate due to an action of the Clifford algebra. It is this degeneracy which makes the triple semi-finite. For details we refer to \cite{AGN3,AGN2}.

\section{Relation to Canonical gravity}

%It turns out that the interaction between the Dirac type operator $D$ and the algebra $\cb_{x_0}$ reproduces the structure of the Poisson bracket of General Relativity when formulated in Ashtekar variables. 

Let us briefly summarize
 canonical gravity formulated in terms of connection variables \cite{Ashtekar:1986yd,Ashtekar:1987gu} and the related formulation in terms of flux and holonomy loop variables, used in Loop Quantum Gravity \cite{AL1}. Let $M$ be a hyperbolic space-time and consider a foliation $M=\Sigma\times \mathbb{R}$, where $\Sigma$ is a spatial 3-dimensional hyper surface. The Ashtekar variables consist of a complexified $SU(2)$ connection $A_n(x)$ and a densitized inverse triad field $E_a^m=\sqrt{g} e_a^m$ on $\Sigma$. Here $n,m,l ...$ and $a,b,c ...$ denote curved and flat spatial indices; $g$ is the determinant of the metric on $\Sigma$ and $e_a^m$ is the inverse dreibein field on $\S$. The Poisson bracket between the connection and triad variables reads
$$
\{A_n^a(x),E_b^m\} = \k\delta^a_b\d_n^m \d^3(x-y)\;,
$$
where $\k$ is the gravitational constant. In addition to this there is a set of three constraints, the Hamilton,  Diffeomorphism and Gauss constraints.%\footnote{These constraints have this form when $A_m(x)$ is complex or when one considers the Euclidean case.}
%\begin{equation}
%\e^{ab}_{\;\;c}  E^n_a E^m_b F^c_{mn}=0\;,\quad E^m_a F_{mn}^a =0\;,\quad \pa_m E^m_a + \e_ac^{\;\;b} A_m^b E^m_c=0\;.
%\label{theconstraints}
%\end{equation}
%Here $F$ is the field strength of the connection $A$.
%

The formulation of canonical gravity in terms of connection variables permits a shift to loop variables, which are taken as the holonomy transform  
\[
h_l(A)  = \cp \mbox{exp}\int_l A_m dx^m\;,
\]
along a loop $l$ in $\Sigma$, %To define a conjugate variable to $h_l(A)$ let $dF_a$ be the flux of the triad field $\bar{E}_a^m$ corresponding to an infinitesimal area element of the spatial manifold $\Sigma$, which can be written
%\begin{equation}
%dF_a = \e_{mnp} \bar{E}^m_a dx^n\wedge dx^p \;.
%\nonumber%\label{infinitesimal}
%\end{equation}
and flux variables, which are the flux of $E_a^m$ associated to a surface $S$ in $\Sigma$
\[
F^{S}_a= \int_S   \e_{mnp} E^m_a dx^n\wedge dx^p  \;.
\]
Let $l=l_1\cdot l_2$ be a line segment in $\Sigma$ which intersect $S$ at the point $l_1\cap l_2$. The Poisson bracket between the flux and holonomy variables read
\begin{equation}
\{ h_l, F^S_a \} =  \pm \kappa h_{l_1}\t_a h_{l_2}\;.
\label{Poisson}
\end{equation}
where $\t$ denote the generators of the Lie algebra of $SU(2)$ and the sign
depends on the intersection between $S$ and $l$.\\

In \cite{AGNP1} we have shown that the commutator between the operator $D$ and elements of the loop algebra $\cb_{x_0}$ reproduces the structure of the bracket (\ref{Poisson}). It turns out that the derivations $\cl_{{\bf e}_i^a}$ corresponds to infinitesimal flux operators located at the vertices in the lattice. Thus, the Dirac operator $D$ can be understood as an infinite sum of all flux operators in the infinite lattice. In this sense  one can see the Hilbert space $\ch$ as a kinematical Hilbert space since it carries a representation of the quantized Poisson structure. Notice, however, that $\ch$ does not carry an action of the diffeomorphism group. \\

%In the following we consider first a real $SU(2)$ connection. In section ... we show how the spectral triple construction can incorporate a complex connection by doubling the Hilbert space.

\section{Coherent states in $\ch$}

We here briefly recall part of the properties of the semi-classical states constructed in \cite{AGNP1}. This construction uses results of Hall \cite{H1,H2}, and is inspired by the articles \cite{BT1,TW,BT2}.

Let $E,A$ be a point in the classical phase space. The semi-classical state $\phi_n^t \in L^2(\overline{\ca})$ with respect to this point have the properties:
\begin{enumerate}
\item For any path $p \in \Gamma_n$ and any $w\in M_2(\bbC )$ \label{hol}
$$\lim_{t \to 0}\langle \phi_n^t \otimes w,h_p \phi_n^t\otimes w\rangle= \langle w,Hol(p,A)w\rangle\;.$$
This in particular means that the expectation value in the limit $n\to \infty$ on a path in $\cup_n\Gamma_n$ is just the holonomy of the connection $A$ along the path.
\item \label{efelt} For an edge $l_i\in \Gamma_n\setminus\Gamma_{n-1}$ in direction $j$
$$\lim_{t \to 0}\langle \phi_n^t,-t i \mathcal{L}_{{\bf e}^a_i}\phi_n^t\rangle =2^{-2n}E_a^j (x)\;,$$  
 where $x$ denotes the right end point of $l_i$. If $l_i \notin \Gamma_n \setminus \Gamma_{n-1}$ the corresponding expectation value will be zero.
\item $\| \phi_n^t\|=1$\;.
\end{enumerate}
The properties \ref{hol}. and \ref{efelt}. also hold for polynomial function on $T^*SU(2)$ %, i.e.
%\begin{eqnarray*}
%\lefteqn{\lim_{t \to 0}\langle \phi_n^t \otimes w,P(h_l,t i \mathcal{L}^e_1,t i \mathcal{L}^e_2 ,it \mathcal{L}^e_3) \phi_n^t\otimes w\rangle}\\
%&=& \langle w,P(Hol(l,A),E_1^i, E_2^i,E_3^i),w\rangle,
%\end{eqnarray*}
and for more general functions in $T^*SU(2)$.
Property \ref{hol}. is a consequence of the peakedness of $\phi_n^t$ around $Hol (p,A)$. In particular when the edge becomes small, $\phi_n^t$ is centered around "$1+\epsilon A$". %Since left and round invariant vector fields coincide in the identity on the group, for small edges we have
%  $$\lim_{t \to 0}\langle \phi_n^t,t i \mathcal{R}^e_a\phi_n^t\rangle \sim 2^{-2n}E_a^i (v_e).$$

%\section{Local gauge invariance, spinor fields and the Dirac Hamiltonian}

\section{Spinor fields and the choice of base-point}

As already stated, the algebra of holonomy loops comes with a dependency on the choice of base-point. If a loop $L_0$ in $\G_n$ based in $x_0$ is related to another loop $L_1$ based in $x_1$ through a unitary transformation
$$
L_1 = \cu_p L_0 \cu^\ast_p\;,
$$
where $\cu_p=\cu_{i_1}\cu_{i_2}\ldots\cu_{i_n}$ is the parallel transport along a path $$p=\{l_{i_1},l_{i_2},\ldots,l_{i_n}\}$$ from $x_0$ to $x_1$ ($\cu_{l_i}=g_i$), then the action of these two loops in associated Hilbert spaces is clearly not identical: the parallel transports $\cu_p$ and $\cu^*_p$ do not cancel. If we were considering an algebra of traced loops, as is the case in Loop Quantum Gravity, then the parallel transports would vanish. Also, if we were considering a state for which the matrix factor was a multiple of the identity the parallel transports would vanish too. But in general they do not.

The purpose of this section is to remove this dependency on the base-point. 
%Consider again the form of the Hilbert space (\ref{Hilbert}) at the level $n$ of subdivision. %, which has the form
%$$
%\ch = L^2(G^{n(\G_n)},Cl(T^\ast G^{n(\G_n)})\otimes M_2(\mathbb{C}))
%$$
%Here, the factor $M_2(\mathbb{C})$ is a 2-by-2 matrix factor which was added to carry a representation of the algebra $\cb_{x_0}$. This algebra is generated by loops based in the base-point $x_0$.
Denote by $\cb^n_{x}$ be the algebra of loops in $\G_n$ based in the vertex $x$. The relationship between $\cb^n_{x_0}$ and $\cb^n_{x_1}$ is given by
$$
\cb^n_{x_0} = \cu_p \cb^n_{x_1} \cu^\ast_p\;.
$$
If we want to eliminate the presence of a preferred base-point we need to incorporate also the shifted algebra $\cb_{x_1}$ in the construction. 
To do this we introduce the operator 
$$
\tilde{\cu}_i =   \mathrm{i}{\bf e}_i^a ( g_i \otimes\beta_i^a )\;,
$$
where $\beta^a_i$ is an arbitrary, self-adjoint matrix associated to the $i$'th edge satisfying 
$$
\sum_a \vert \beta^a_i\vert^2 =1\;,\quad (\beta_i^a)^* = -\beta_i^a
$$
and where $g_i\otimes \beta_i^a$ refers to left and right actions in $\ch_n$ in the sense that $g_i$ acts identically  from the left on the two copies of $M_2(\mathbb{C})$ and $\beta_i^a$ acts from the right, i.e. $\beta_i^a$ is a four-by-four matrix. Note that $\tilde{\cu}_i$ is not unitary.
%\begin{equation}
%\tilde{\cu}_i \tilde{\cu}^\ast_i  = 1 + [{\bf e}_i^a ,{\bf e}_i^b ] (\beta_i^a (\beta_i^b)^* - \beta_i^b (\beta_i^a )^*) \;.
%\label{tilde}
%\end{equation}
Also, we introduce the operators
$$
\tilde{\cu}_p  =  %(\mathrm{i}{\bf e}_{i_1}^a \sigma^a g_{i_1})  (\mathrm{i}{\bf e}_{i_2}^a \sigma^a g_{i_2}) \cdot \ldots\cdot  (\mathrm{i}{\bf e}_{i_n}^a \sigma^a g_{i_n}) 
\tilde{\cu}_{i_1}\tilde{\cu}_{i_2}\cdot\ldots\cdot\tilde{\cu}_{i_n}
$$
associated to the path $p$
and note that these operators are mutually orthogonal
\begin{equation}
\langle \tilde{\cu}_p \vert \tilde{\cu}_{p'}\rangle = 
\left\{
\begin{array}{cl}
1 & \mbox{if}\quad p=p'\\
0 & \mbox{if}\quad p\not=p'
\end{array}
\right.
\label{nn}
\end{equation}
due to their dependency on the Clifford algebra. The inner product in (\ref{nn}) is the inner product in $\ch$ which involves the trace over the Clifford algebra. Next, let $\{\psi(x_i)\}$ be a family of matrices in $M_2(\mathbb{C})\oplus M_2(\mathbb{C})$ associated to vertices in $\G_n$ and transform them to
\begin{equation}
\tilde{\cu}_p\psi(x_i)\;.
\label{shifted}
\end{equation}
Consider again the loop $L_0$ based in $\cb_{x_0}$ and two corresponding loops $L_1$ in $\cb_{x_1}$ and $L_2$ in $\cb_{x_2}$ shifted by paths $p_1$ and $p_2$
$$
L_1 = \cu_{p_1}L_0\cu_{p_1}^\ast\;,\quad L_2 = \cu_{p_2}L_0\cu_{p_2}^\ast\;.
$$
One easily checks that
\begin{eqnarray}
\langle \tilde{\cu}_{p_1}\psi(x_1) + \tilde{\cu}_{p_2}\psi(x_2)  \vert L_0 \vert \tilde{\cu}_{p_1}\psi(x_1) + \tilde{\cu}_{p_2}\psi(x_2) \rangle &=& \nn\\
 &&\hspace{-3cm}   \langle \psi(x_1) \vert L_1\vert \psi(x_1)\rangle + \langle \psi(x_2) \vert L_2\vert \psi(x_2)\rangle \;, 
 \nn
\end{eqnarray}
which shows that the sum $ \tilde{\cu}_{p_1}\psi(x_1) + \tilde{\cu}_{p_2}\psi(x_2) $ carries a representation of both algebras $\cb_{x_1}$ and $\cb_{x_2}$ simultaneously. Therefore, to eliminate the choice of base-point all we need to do is to add up all the factors (\ref{shifted}) for all the vertices
\begin{equation}
 \xi_n(\psi)=\frac{1}{\cn} \sum_{i=1} \tilde{\cu}_{p_i} \psi(x_i)
\label{xi}
\end{equation}
to obtain a construction which takes all possible base-points into account simultaneously. 
Clearly, the rhs of (\ref{xi}) depends on  a choice of paths between the base-point $x_0$ and vertices in the sum. In the next section we shall further specify the sum in (\ref{xi}) as well as the normalization $\cn$.

\section{The Dirac Hamiltonian}

\begin{figure}[t]
\begin{center}
\resizebox{!}{4cm}{
 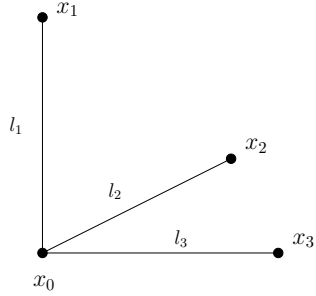}
\end{center}
\caption{\it three edges, $l_1,l_2,l_3$, connected in one vertex $x_0$.}
\end{figure}
We wish to compute the expectation value of the Dirac type operator $D$ on the state (\ref{xi}) in a semi-classical approximation. Let us first consider the system of a single base-point $x_0$ and three edges $l_1,l_2,l_3$ connecting the base-point with vertices $x_1,x_2,x_3$, see figure 2. In this case we let the state (\ref{xi}) involve four terms
$$
\tilde{\cu}_0 \psi(x_0)\;,\quad\tilde{\cu}_1 \psi(x_1)\;,\quad\tilde{\cu}_2 \psi(x_2)\;,\quad\tilde{\cu}_3 \psi(x_3)\;,
$$
where $\tilde{\cu}_0:= \mathds{1}$.
If we calculate the expectation value of $D$ on the state
$$
\Psi^t_{x_0}(\psi)=
\left(\frac{1}{4} \sum_{i=0}^3 \tilde{\cu}_i \psi(x_i)\right)  \phi^t_n\;,
$$ 
where $\phi^t_n$ are the coherent states introduced in section 5, and if we take the limit where the edges $l_1,l_2,l_3$ lie infinitely deep in the inductive system of lattices, then we find
\begin{eqnarray}
\lim_{n\rightarrow\infty}\lim_{t\rightarrow 0}\langle \bar{\Psi}^t_{x_0} \vert t D \vert\Psi^t_{x_0}\rangle &=& \mbox{Tr}\; \bar{\psi}(x_0) \left( E_a^i \nabla_i +\nabla_i E_a^i     \right)\psi(x_0)\beta^a\;,
%\nn\\ &&
% + \mbox{\it zero order terms.}
\label{D}
\end{eqnarray}
where the trace is wrt $SU(2)$ and where we used $g_i\psi(x_i)=(\mathds{1} +\nabla_i)\psi(x_0)$, with $\nabla_i = \pa_i + A_i$, which holds in the limits taken. Further, we took $\beta^a_1=\beta^a_2=\beta^a_3\equiv \beta^a$ and set $a_n=2^{3n}$. If we write $\beta^a= N(x_0)\gamma^a+\mathrm{i}N^a(x_0)\gamma^0$, where $\gamma^a$, $\gamma^0$, $a\in \{1,2,3\}$, are the gamma matrices in the Dirac representation, then (\ref{D}) resembles the expression for the integrand of the Dirac Hamiltonian, with $N(x)$ and $N^a(x)$ being the shift and lapse fields respectively. %Notice, however, that $\psi(x_0)$ takes values in $M_2(\mathbb{C})\oplus M_2(\mathbb{C})$ and that the Dirac matrices act on the right of $\psi$ whereas the connection $A_i(x)$ act on the left. 

%The reason for adding two matrix factors in (\ref{HIL}) is that with only one factor we would not be able to include the shift field $N^a$ in $\beta^a$. In short, it is necessary that both terms in $\beta^a$ are odd with respect to the Clifford algebra in order for (\ref{})

In the general case notice that matrix elements
$$
\langle \tilde{\cu}_p\psi(x_i)  \vert D \vert \tilde{\cu}_{p'}\psi(x_j)  \rangle   
$$
will only give a nonzero result if a relation like
$
\tilde{\cu}_p =  \tilde{\cu}_k\tilde{\cu}_{p'}
$
holds, due to the involved elements in the Clifford algebra. This implies that the vertices $x_i$ and $x_j$ must be only a single edge apart. This is the key mechanism in the following analysis.

Consider the $n$'th level of subdivision and let $\psi(x)$ be smooth field in the 3-manifold $\Sigma$ with values in $M_2(\mathbb{C})\oplus M_2(\mathbb{C})$. Define the state
\begin{equation}
\Psi^t_n(\psi) =\xi_n(\psi)
\phi_n^t\;,
\label{1particlestate}
\end{equation}
%with
%$$
%\Phi_n^t= \prod_{j=1}^{n(\G_n)} \phi^t_{l_i}
%$$
%and 
with $\xi_n(\psi)$  defined in (\ref{xi})
%$$
%\xi'_n(\psi)=\frac{1}{n(\G_n)}  \sum_{i=1}^{n(\G_n)} \cu_{p_i} \psi(x_i)
%$$
%where $V=V(n)$ is the number of vertices at the $n$'th level of subdivision and where $E=E(n)$ is the corresponding number of edges.
where we now specify the sum to run over edges in $ \Gamma_n \setminus \Gamma_{n-1}$ only and where we set the normalization to $\cn=2^{3(n-1)/2}$. This normalization will descent to the Lebesque measure when $n$ approaches infinity. %Also, in (\ref{1particlestate}) we understand $\cu_{p_i}$ to be a unitary transformation corresponding to a path $p_i$ which has the edge $l_i$ at its end. Thus, this state involves a choice of paths.

%The state (\ref{1particlestate}) will be called a one-particle state. 

The expectation value of $D$ on the state $\Psi^t_n(\psi)$ in the  semi-classical large $n$ limit gives
\begin{eqnarray}
\lim_{n\rightarrow\infty} \lim_{t\rightarrow 0}  \langle \bar{\Psi}^t_n \vert  D \vert \Psi^t_n  \rangle 
\nn\\
&&\hspace{-40mm}
=  \int_\Sigma d^3x \bar{ \psi}(x)  (\sqrt{g}  e_a^m\nabla_m +\nabla_m\sqrt{g}  e_a^m )  \psi(x)\left(N(x)\gamma^a + \mathrm{i}N^a(x)\gamma^0   \right) 
%\nonumber\\
%&&\hspace{-40mm}+ \mbox{\it zero order terms.}
   \;.
%\nn
\end{eqnarray}
This expression resembles the Dirac Hamiltonian in 3+1 dimensions, with two important deviations: first, the field $\psi(x)$ takes values in $M_2(\mathbb{C})\oplus M_2(\mathbb{C})$ and thus display a doubling of the degrees of freedom; second, the gamma matrices act from the right on $\psi(x)$.

%The derivation  follows the principles already put forward in the papers \cite{} with the only difference being the exact formulation of the states $\Psi_n^t$, in particular the organization of the elements of the CAR algebra. Clearly, the advance made here is the clear interpretation of the state (\ref{1particlestate}) : where it in \cite{} was engineered to produce the Dirac Hamiltonian it here appears quite natural.

The construction of the states (\ref{1particlestate}) and the emergence of the structure of the Dirac Hamiltonian follows the principles already presented in \cite{Aastrup:2009dy,AGNP1} and therefore most of the commentary given in \cite{Aastrup:2009dy,AGNP1} equally applies here. The key difference is the new naturalness of the state (\ref{1particlestate}).

Note that the failure of $\tilde{\cu}_i$ to be unitary can be traced back to the presence of the shift field: if we wrote $\beta^a = N\gamma^a$ then $\tilde{\cu}_i^a$ could be made unitary by adding a term proportional to ${\bf e}^1_i {\bf e}_i^2 {\bf e}_i^3 $. However, with a non-zero shift field this is no longer possible.

The investigation of the dependence of our construction and interpretation of coherent states on the choice of the foliation will be an important
issue in our future work.\\

%This results raises the question whether there are also natural states which corresponds, in the semi-classical limit, to many-particle states coupled to a gravitational field?\\ %Since the CAR (canonical anticommutation relation) algebra emerges in the continuum limit of the construction (from the infinite dimensional Clifford algebra necessitated by the construction of a Dirac type operator) it seems plausible that a Fock space structure might .

%\section{Discussion}

%The structure of the Dirac Hamiltonian in 3+1 dimensions is shown to emerge from a abstract spectral triple construction over a configuration space of connections. The starting point is a spectral triple, a geometrical construction which captures information about the kinematics of canonical quantum gravity formulated in terms of connection variables. The emergence of the structure of the Dirac Hamiltonian follows from the observation, that the algebra of loops comes with a dependency of a choice of base-point. This is an immediate consequence of considering the noncommutative (untraced) algebra of holonomy loops. It is the elimination of this dependency on the base-point which leads to fermionic fields and, in the semi-classical approximation, to the structure of the Dirac Hamiltonian. The latter emerges from the expectation value of the abstract Dirac type operator on states which combines semi-classical approximation with the elimination of the choice of base-point.

\noindent{\bf\large Acknowledgements}\\

\noindent %d

We would like to thank Ryszard Nest for numerous discussions.
%J.A. and M.P. were supported by the SFB 478 grant  "Geometrische Strukturen in der Mathematik" of the Deutsche Forschungsgemeinschaft.

\end{document}